\documentclass[twocolumn,showpacs,preprintnumbers,amsmath,amssymb]{revtex4}
\usepackage{graphicx}
\usepackage{epsfig}
\usepackage{dcolumn}
\usepackage{bm}

\newcommand{\rar}{\rightarrow}
\newcommand{\jpsi}{J/\psi}
\newcommand{\psip}{\psi(2S)}
\newcommand{\psipto}{\psi(2S)\rightarrow}
\newcommand{\jpsito}{J/\psi\rightarrow}
\newcommand{\pipi}{\pi^+\pi^-}
\newcommand{\pipipi}{\pi^+\pi^-\pi^0}
\newcommand{\gamg}{\gamma\gamma}
\newcommand{\psiptoomegapi}{\psi(2S)\rightarrow\omega\pi^{0}}

\newcommand{\wpi}{\omega\pi^0}

\newcommand{\eetowpi}{e^{+}e^{-}\rightarrow\omega\pi^{0}}
\newcommand{\ee}{e^{+}e^{-}}
\newcommand{\eeto}{e^{+}e^{-}\rightarrow}

\newcommand{\GeV}{~\hbox{GeV}}
\newcommand{\GC}{~\hbox{GeV}/c^2}
\newcommand{\etap}{\eta^{\prime}}
\newcommand{\vp}{\wpi,~\rho\eta,~\rho\etap}

\def\Journal#1&#2&#3(#4){#1 {\bf #2}, #3 (#4)}

\def\NPB{Nucl.  Phys.  {\bf B}}
\def\PLB{Phys.  Lett.  {\bf B}}

\def\PRD{Phys.  Rev.  {\bf D}}

\def\etal{{\it et al.}}
\def\calB{{\cal B}}
\def\bec{\begin{center}}
\def\eec{\end{center}}

\begin{document}

\title{\boldmath Measurement of the final states $\omega
 \pi^0$, $\rho \eta$, and $\rho \eta^{'}$ from $\psip$ electromagnetic
 decays and $\ee$ annihilations}
\author{
M.~Ablikim$^{1}$,        J.~Z.~Bai$^{1}$,           Y.~Ban$^{11}$,
J.~G.~Bian$^{1}$,        X.~Cai$^{1}$,              J.~F.~Chang$^{1}$,
H.~F.~Chen$^{17}$,       H.~S.~Chen$^{1}$,          H.~X.~Chen$^{1}$,
J.~C.~Chen$^{1}$
% ...
\\(BES Collaboration)\\ 
\vspace{0.2cm}
$^{1}$ Institute of High Energy Physics, Beijing 100049, People's Republic of China\\
% ...
$^{11}$ Peking University, Beijing 100871, People's Republic of China\\
% ...
$^{17}$ University of Science and Technology of China, Hefei 230026, People's Republic of China\\
% ...
}

\begin{abstract}
  Cross sections and form factors for $\ee \to \wpi$, $\rho\eta$, and
  $\rho\etap$ at
  center of mass energies of 3.650, 3.686, and 3.773 GeV are measured
  using data samples collected with the BESII detector at the BEPC.
  Also, the branching fractions of $\psi(2S) \rar \wpi$, $\rho\eta$,
  and $\rho\etap$ are determined to be
  $(1.87^{+0.68}_{-0.62}\pm0.28)\times 10^{-5}$,
  $(1.78^{+0.67}_{-0.62}\pm0.17)\times 10^{-5}$, and
  $(1.87^{+1.64}_{-1.11}\pm0.33)\times10^{-5}$, respectively.
\end{abstract}
\maketitle

\section{Introduction}
Form factors of the electromagnetic processes $\psipto\wpi$
($\rho\eta,~\rho\etap$) provide information on the strength of the
electromagnetic amplitude in $\psip \rar 1^-0^-$ decays. Such
information is indispensable  to separate the strong interaction
amplitude for other $\psip \rar 1^-0^-$
decay modes, and
even more importantly, to determine the relative phase between the
electromagnetic and strong amplitudes~\cite{phase}. This phase has
been found to be orthogonal in virtually all $J/\psi$ decays, such as
$1^+0^-$~\cite{suzuki}, $1^-0^-$~\cite{jousset, coffman},
$0^-0^-$~\cite{suzuki1, jdecay}, $1^-1^-$~\cite{jdecay}, and
$N\bar{N}$~\cite{nnbar} modes. It is suggested that this orthogonality
also exists in $\psip$ decays and is universal~\cite{plb574}.

The Born-order cross section for the electromagnetic  process 
$\eeto V + P$ $(\vp)$ in  
continuum production can be expressed as~\cite{wangp}
\begin{equation}\label{Fzh}
\sigma_{Born}(s) = \frac{4\pi\alpha^2}{s^{3/2}}\cdot
|\mathcal{F}_{VP}(s)|^2 \cdot \mathcal{P}_{VP}(s) ,
\end{equation}
where $\mathcal{P}_{VP}(s)= \frac{1}{3} q^3_{VP}$,
$q_{VP}$ is the momentum of either the Vector or Pseudoscalar meson in the VP
decay, and the form factor $|\mathcal{F}_{VP}(s)|$ is an $s$ dependent
variable~\cite{chernyak, Gerard}. 

By measuring the cross section of $\eeto VP$ in continuum production
and correcting for the effect of initial state radiation, the form
factor $\mathcal{F}_{VP}(s)$ can be determined. In contrast, for
$\eeto VP$ at the $\psip$ peak, the total cross section includes
contributions from both resonance production and decay and the
continuum process, although the interference between them can be
neglected~\cite{wangp}. By separating the one-photon annihilation and
$\psip$ resonance contributions, both the form factor at resonance and
the branching fractions of $\psipto VP$ can be obtained. The branching
fractions of $\psipto\rho\eta$ and $\rho\etap$ provide useful
information on the quark content of $\eta$ and $\etap$
mesons~\cite{jdecay}.

In this paper, we report measurements for the cross sections and form 
factors of $\wpi$, $\rho\eta$, and $\rho\etap$ at center of mass energies 
3.650, 3.686, and 3.773 GeV, and the
branching fractions of $\psipto$ $\wpi$, $\rho\eta$, and $\rho\etap$ at
3.686 GeV.

\section{THE BESII DETECTOR}
The Beijing Spectrometer (BESII) is a conventional cylindrical
magnetic detector that is described in detail in Ref.~\cite{BES-II}.
A 12-layer Vertex Chamber (VC) surrounding the beryllium beam pipe
provides input to the event trigger, as well as coordinate
information.  A forty-layer main drift chamber (MDC) located just
outside the VC yields precise measurements of charged particle
trajectories with a solid angle coverage of $85\%$ of $4\pi$; it also
provides ionization energy loss ($dE/dx$) measurements which are used
for particle identification.  Momentum resolution of
$1.7\%\sqrt{1+p^2}$ (p in GeV/$c$) and $dE/dx$ resolution for hadron
tracks of $\sim8\%$ are obtained.  An array of 48 scintillation
counters surrounding the MDC measures the time of flight (TOF) of
charged particles with a resolution of about 200 ps for hadrons.
Outside the TOF counters, a 12 radiation length, lead-gas barrel
shower counter (BSC), operating in limited streamer mode, measures the
energies of electrons and photons over $80\%$ of the total solid angle
with an energy resolution of $\sigma_E/E=0.22/\sqrt{E}$ ($E$ in GeV).  A
solenoidal magnet outside the BSC provides a 0.4 T magnetic field in
the central tracking region of the detector. Three double-layer muon
counters instrument the magnet flux return and serve to identify muons
with momentum greater than 500 MeV/$c$. They cover $68\%$ of the total
solid angle.

In this analysis, a GEANT3 based Monte Carlo package (SIMBES) with detailed
consideration of the detector performance (such as dead electronic channels)
 is used. The consistency between data and Monte Carlo (MC) has been carefully 
checked in many high purity physics channels, and the agreement is reasonable.
The generators {\bf EE2VP}~\cite{ee2wpi} and {\bf HOWL-VP}~\cite{howl-vp},
together with SIMBES, are used
to determine the detection efficiencies for the one photon annihilation
and resonance decay processes.

\section{EVENT SELECTION} 
The data samples used for this analysis consist of $14.0\times10^6~(1\pm4\%)$
$\psi(2S)$ events~\cite{Npsip},  $6.42~(1\pm4\%)$ pb$^{-1}$ of
continuum data at $\sqrt{s}=3.650\GeV$~\cite{Lcont}, and
17.3~(1$\pm$3\%)~pb$^{-1}$ at the $\psi(3770)$~\cite{Lpsipp}. 
The channels studied are $\wpi$,
 $\rho\eta$, and $\rho\etap$,  where $\omega$ decays to $\pi^+\pi^-\pi^0$, 
$\rho$ to $\pipi$, $\eta$ to $\gamg$, 
and $\etap$ to $\eta\pi^+\pi^-$.

A neutral cluster is considered to be a photon candidate if it is
 located within the BSC fiducial region, the energy deposited in the
 BSC is greater than 50 MeV, the first hit appears in the first 6
 radiation lengths, and the angle between the cluster development
 direction in the BSC and the photon emission direction from the beam
 interaction point (IP) is less than $37^\circ$. For the $\rho\eta$
 channel, tighter requirements are applied: the energy deposited in
 the BSC must be greater than 100 MeV, and the angle in the $x y$ plane
 (perpendicular to beam direction) between the neutral cluster and the
 nearest charged track must be greater than $15^\circ$.

 Each charged track is required  to be well fit by a three-dimensional
helix, to originate from the IP region, $V_{xy}=\sqrt{V_x^2+V_y^2}<2$ cm and
$|V_z|<20$ cm, and to have a polar angle $|\cos\theta|<0.8$. Here $V_x$, 
$V_y$, and $V_z$ are the x, y, and z coordinates of the point of the closest
approach of the track to the beam axis.

The TOF and $dE/dx$ measurements for each charged track are used to
calculate $\chi^2_{PID}(i)$ values and the corresponding confidence levels
$Prob_{PID}(i)$
for the hypotheses that
a track is a pion, kaon, or proton,
where $i$ ($i=\pi/K/p$) is the particle type.
For events involving the final state $\wpi$ and $\rho\etap$,  at least
half of charged pion candidates in each event are required to have 
$Prob_{PID}(\pi)$ larger than 0.01, while for events 
involving the final state $\rho\eta$, both charged pion candidates must satisfy this requirement.

\subsection{\boldmath $\wpi$ channel}
For this channel, the events are  required to have two good charged tracks
with net charge zero and four or five photon candidates.
A four constraint (4C) kinematic fit under the hypothesis 
$\ee\rar\pi^+\pi^-\gamg\gamg$ is performed. If the number of selected photons is
larger than four, the fit is repeated using all possible combinations of
 photons, and the one with the smallest $\chi^{2}$
is chosen. The confidence level of the 4C fit is required to be 
larger than 0.01. In addition, we require that 
$\chi^2_{com}$~\cite{Chisquare} for the assignment 
$\ee\rightarrow\pi^+\pi^-\gamg\gamg$ must be smaller than that for
$\ee\rightarrow K^+K^-\gamg\gamg$ in order to suppress possible
$K^+K^- 4\gamma$ background.

Among the four selected photons, there are three possible combinations 
to compose two $\pi^0$s : 
($\gamma_{1}\gamma_{2}$, $\gamma_{3}\gamma_{4}$),
($\gamma_{1}\gamma_{3}$, $\gamma_{2}\gamma_{4}$), and 
($\gamma_{1}\gamma_{4}$, $\gamma_{2}\gamma_{3}$).
Events where one and only one combination have both $\pi^0$ 
candidates satisfying $|M_{\gamg} -0.135|<0.05 \GC$ are kept for further
analysis. 

\begin{figure}[hbt]
\includegraphics[width=0.45\textwidth]{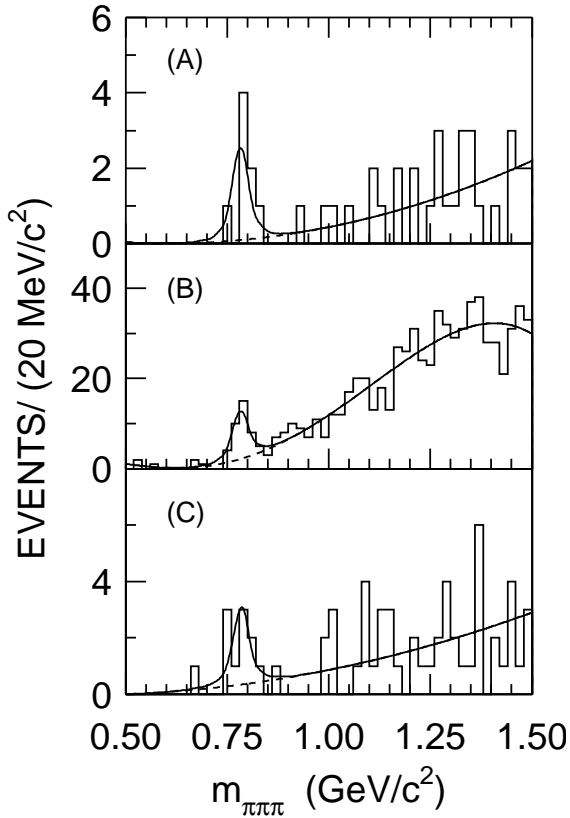}
\caption{\label{Fresult1}The $\pipi\pi^0_{L}$ invariant mass distributions for
$\ee\rar\pipi\pi^0_{L}\pi^0_{H}$ candidate events at $\sqrt{s}=$ 
(A) 3.650, (B) 3.686, and (C) 3.773 GeV.  $\pi^0_{L}$ and
  $\pi^0_{H}$ are the low and high energy pions. 
The curves show the fit
described in the text. }
\end{figure}

Using the above selection,  $\pi^+\pi^-\pi^0\pi^0$ candidate events are 
obtained. Fig.~\ref{Fresult1} shows the $\pipi\pi^0_{L}$ invariant mass 
distributions, where $\pi^0_{L}$ is the lower energy $\pi^0$ from the 
two $\pi^0$ candidates. Clear $\omega$ signals are seen in all three data 
samples at $\sqrt{s}=$3.650, 3.686, and 3.773 GeV.

The $\pipi\pi^0_{L}$ invariant mass distribution is fit with a shape
for the signal determined by MC simulation plus a polynomial
background, and $7.3^{+3.3}_{-2.7}$, $31.2^{+7.7}_{-6.9}$, and
$8.6^{+4.0}_{-3.3}$ events are observed in the data samples at
$\sqrt{s}=$3.650, 3.686, and 3.773 GeV, with $3.9\sigma$, $5.6\sigma$,
and $3.2\sigma$ statistical significance~\cite{significance},
respectively.

\subsection{\boldmath $\rho\eta$ channel}
\begin{figure}[hbt]
\includegraphics[width=0.45\textwidth]{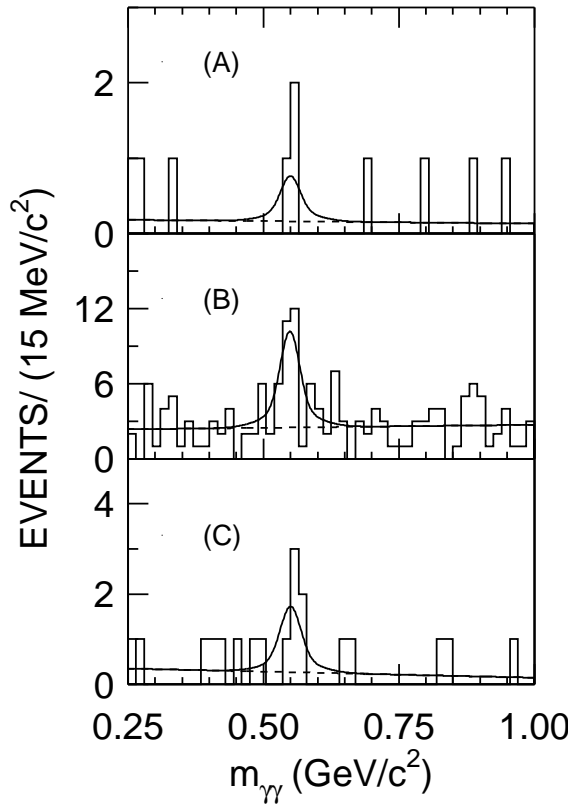}
\caption{\label{Fresult2}The $\gamg$ invariant mass distributions for
$\ee\rar\rho\gamg$ candidate events at  $\sqrt{s}=$ 
(A) 3.650, (B) 3.686,  and (C)  3.773 GeV.
The curves show the fit
described in the text.}
\end{figure}

For this channel, the events are required to have
 two good charged tracks with net charge zero and two photon candidates.
A 4C kinematic fit to the hypothesis $\ee\rar\pi^+\pi^- \gamg $ is performed, 
and its confidence level is required to be larger than 0.01 and larger than that 
 for the assignment $K^+K^- \gamg$  to suppress 
possible $K^+K^- \gamg$ background. The energy of each 
photon is required to be less than 1.7 GeV to reject backgrounds from
$\psipto\gamma\etap$. An additional requirement
$|m_{\pipi}-0.776|<0.15 \GC$ is applied to further suppress background 
from non-$\rho$ decay. 
Fig.~\ref{Fresult2} shows the $\gamg$ invariant mass distributions for 
$\rho\gamg$ candidate events. Clear $\eta$ signals are seen in all 
three data samples at $\sqrt{s}=$3.650, 3.686, and 3.773 GeV.

The $m_{\gamg}$ invariant mass spectrum is fitted with a shape for the $\eta$ signal determined by MC
simulation plus a polynomial background, and $2.3^{+2.1}_{-1.4}$, 
$29.2^{+7.5}_{-6.8}$, and $5.8^{+3.3}_{-2.6}$ events 
are observed in the data samples at $\sqrt{s}=$3.650, 3.686, 
and 3.773 GeV, respectively.
The statistical significances are $1.9\sigma$, $5.5\sigma$, and 
$3.3\sigma$,  respectively.

\subsection{\boldmath $\rho\etap$ channel}
\begin{figure}[hbt]
\includegraphics[width=0.45\textwidth]{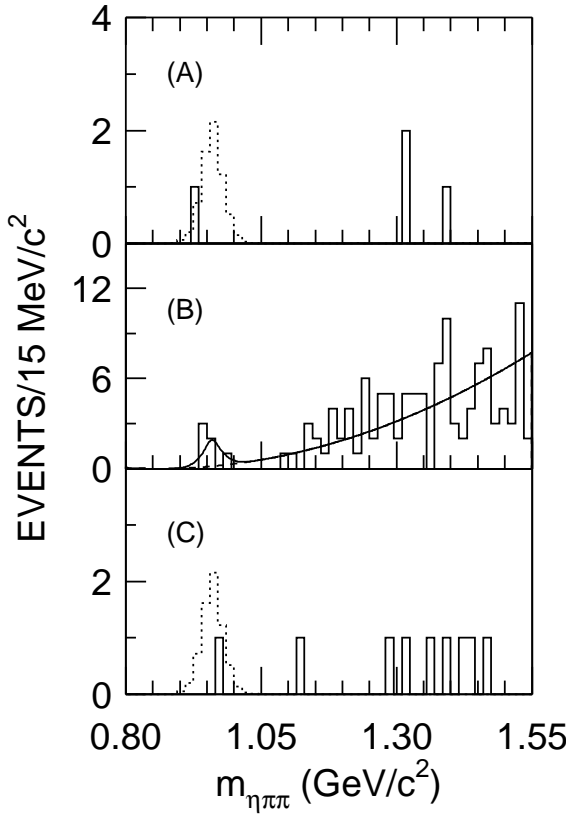}
\caption{\label{Fresult3}The $\eta\pipi$ invariant mass distributions
for $\ee\rar\rho\pipi\eta$ candidate events (solid histogram) and
$\ee\rar\rho\etap$ Monte Carlo simulation (dotted histogram) with
arbitrary normalization for  $\sqrt{s}=$  (A) 3.650, (B) 3.686,  and
(C) 3.773 GeV. The curve in (B) shows the fit described in the text.}
\end{figure}

Here, events with four good charged tracks with net charge zero and
two or three photon candidates are selected. A 4C kinematic fit is
performed for the hypothesis $\pipi\pipi \gamg$. If the number of
selected photons is larger than 2, the fit is repeated using all
combinations of photons, and the one with the smallest $\chi^{2}$
is chosen. The confidence level of the 4C fit is required to be larger than
0.01 and larger than that for $K^+K^-\pipi \gamg$.
The two photons are required to come from $\eta$ decay
($|m_{\gamg}-0.548|<0.05 \GC$). Background from $\psipto\pipi J/\psi$
is rejected with the requirement that the mass recoiling from any
$\pipi$ pair satisfies $|m^{\pi^{+}\pi^{-}}_{recoil}-3.1|>0.05\GeV/c^2$.  
In order to suppress background from non-$\rho$ decay, an
additional requirement $|m_{\pi^+\pi^-}-0.776|<0.15 \GC$ is applied,
where $m_{\pi^+\pi^-}$ runs over all possible $\pipi$ pairs.
Fig.~\ref{Fresult3} shows the $\eta\pi^+\pi^-$ invariant mass
distributions for $\rho\pipi\eta$ candidate
events. $5.4^{+3.3}_{-2.2}$ events are obtained in the data sample at
$\sqrt{s}=3.686\GeV$ by fitting the $\eta\pi^+\pi^-$ invariant mass
spectrum with an $\etap$ shape obtained by MC simulation plus a
polynomial for background.  The statistical significance for the
$\etap$ signal is 3.1$\sigma$. Only 1 event is observed in each of
the two data samples at $\sqrt{s}=$3.650 and 3.773 GeV, and the respective
background is zero and 0.64 events as estimated from sidebands; the
corresponding upper limit for the observed $\rho\etap$ event is
calculated to be 4.4 and 3.9 events, respectively, with the scheme of
J. Conrad \etal~\cite{upperlimit}.

\section{Separation of continuum and resonance events}
At the $\psip$ peak, the cross section is from resonance decay,
one-photon annihilation, and their interference, all electromagnetic
processes.  The observed events are separated for the $\psip$ decay
and one-photon annihilation processes using the scheme proposed by
P. Wang \etal~\cite{wangp}, considering the detection-efficiency
difference between the resonance decay and one-photon annihilation,
and neglecting the tiny interference (see
Table~\ref{Tffwpi}), that is, 
$$N^{obs}=\mathcal{L}\cdot(\sigma^{\psip}\cdot \epsilon^{\psip} +
\sigma^{Cont.}\cdot \epsilon^{Cont.}),$$
where $\mathcal{L}$ is the integrated luminosity, $\sigma$ the 
cross section, and $\epsilon$ the detection efficiency.  
Table~\ref{Tffwpi} lists detailed information on
the separation of the signal events observed in the data sample at
$\sqrt{s}=3.686 \GeV$.

For the data sample at $\sqrt{s}=3.650\GeV$, the fraction 
of cross section from the $\psip$ resonance is negligible ($<1.1\times 10^{-4}$) 
compared to that of the continuum, as estimated using the 
scheme proposed by P. Wang \etal~\cite{wangp}. Therefore we attribute
all observed events to one-photon annihilation.

For the data sample at $\sqrt{s}=3.773 \GeV$ ,
the resonance part of the cross section for
$\ee\rar\psi(3770)\rar VP~(\vp)$, at Born order, is
\begin{equation}\label{psippwpi}
\sigma_{Born}(s) = \frac{12\pi\Gamma_{ee}\Gamma_{f}}
                {(s-M^2)^2+\Gamma_{t}^2M^2},
\end{equation}
where $\Gamma_{f}$ is the partial width to the final state $f$ ($\vp$) and
is related to $\Gamma_{ee}$ and the corresponding 
form factor is~\cite{wangp}:
$$\Gamma_{f} = \frac{\Gamma_{ee} q^3_{f}}{m_{\psi(3770)}}
|\mathcal{F}_{f}(m^2_{\psi(3770)})|^2.$$
Since $\Gamma_{t}$ and $\Gamma_{ee}$ 
of $\psi(3770)$ are $23.6\pm2.7 ~\hbox{MeV}/c^2$ 
and $0.26\pm0.04 ~\hbox{keV}/c^2$,
respectively~\cite{PDG}, 
the fraction of the cross section from $\psi(3770)$ resonance
production is 
negligible ($\approx 2.1\times 10^{-5}$) 
compared to that from one-photon annihilation. The 
$\psip$ tail at $\sqrt{s}=3.773 \GeV$ contributes no more than 2.0\% for
the observed events in these final states.
Therefore, the observed events
are taken to be totally from one-photon annihilation.

\begin{table*}[htpb] 
\caption{\label{Tffwpi} Cross section and event fractions for the $\psip$
  production and 
decay and one-photon annihilation processes for $\wpi$, $\rho\eta$,
  and $\rho\etap$  
final states at $\sqrt{s}=3.686 \GeV$. 
$\epsilon$ is the detection efficiency determined from MC simulation.} 
\begin{ruledtabular}
\begin{tabular}{l|cc|cc|cc|c } 
        & \multicolumn{2}{c|}{$\wpi$} & \multicolumn{2}{c|}{$\rho\eta$} &
                \multicolumn{2}{c|}{$\rho\etap$} & \\ \hline 
 & $\psip$&  Cont. & $\psip$&  Cont. & $\psip$&  Cont. &  Int. \\  \hline 
$\sigma$ frac. (\%)  & 40.6 & 60.7 & 40.6 &  60.7 & 40.8   &  60.5 & -1.3 \\ \hline
$\epsilon$ (\%)& 5.97 & 4.98 & 13.43 & 10.89  & 5.48 &  4.43 &  -   \\ \hline
$N^{obs}$ frac. (\%) & 44.5 & 55.5 & 45.2  &  54.8  & 45.5 &  54.5  &  - \\ 
\hline
$N^{obs}$     & $13.9^{+5.1}_{-4.6}$ & $17.3^{+5.7}_{-5.1}$ 
              & $13.2^{+5.0}_{-4.6}$ & $16.0^{+5.6}_{-5.0}$
              & $2.5^{+2.2}_{-1.5}$  & $2.9^{+2.4}_{-1.6}$  & - \\
\end{tabular} 
\end{ruledtabular}
\end{table*}

\section{Systematic error}
Many sources of systematic error are considered. 
Systematic errors
associated with the efficiency are determined by comparing $J/\psi$
and $\psi(2S)$ data with Monte Carlo simulations for very clean 
decay channels, such as
$\jpsi\rar\rho\pi$, $\wpi$ and $\psi(2S) \rar \pi^+ \pi^- J/\psi$, which
allows the determination of systematic errors associated with
the MDC tracking efficiency, kinematic fitting and photon selection 
efficiencies, etc.~\cite{systematics}

To investigate possible background channels, we utilize three Monte Carlo
samples : I) continuum channels from $u$, $d$, and $s$ quark 
fragmentation generated with JETSET7.4~\cite{jetset};  II) 
$\psipto {anything}$ generated with LUND-charm generator~\cite{chenjc};
and III) $D\bar{D}$ pairs generated at $\sqrt{s}=3.773 \GeV$. From these 
MC samples, we find that the background contaminations 
are negligible in the three data samples, 
except for the $\wpi$ channel at $\sqrt{s}=3.686 \GeV$.
Events from $\psip\to b_1^0\pi^0$  and cascade decays 
$\psipto anything + \jpsi$,  $\jpsi\to\rho^0\pi^0$  can produce background
in the $M_{\omega}$ mass region. The resulting estimate of 
contamination from these sources is $N_{bkg}=1.6\pm0.7$, and the 
corresponding systematic error is (5.1$\pm$2.3)\%. We take 7\% as a 
conservative estimation and treat it as one source of systematic 
error for the $\wpi$ channel at $\sqrt{s}=3.686 \GeV$.

Different $s$ dependences of the form factor ($\frac{1}{s}$ versus 
$\frac{1}{s^2}$)
result in changes of 2.5\%, which is taken as one of the
systematic errors. 
The uncertainties of the generators, background shapes, and 
luminosities
are also included.  Table ~\ref{TSystematicError} lists 
all the sources of systematic errors, and the total systematic error 
is taken as the sum of the individual terms added in quadrature.
\begin{table}[h]
\caption{\label{TSystematicError} 
Summary of systematic errors ($\%$). D1, D2, and D3 represent 
the data samples at $\sqrt{s}=3.65,$ 3.686, and 3.773 GeV, respectively.}
\begin{ruledtabular}
\begin{tabular}{l|ccc|ccc|ccc}
     & \multicolumn{3}{c|}{$\wpi$} & \multicolumn{3}{c|}{$\rho\eta$}
      & \multicolumn{3}{c}{$\rho\etap$} \\ \hline
Sample &D1&D2&D3  &D1&D2&D3  &D1&D2&D3 \\  \hline
Tracking      &4.0&4.0&4.0   &4.0&4.0&4.0   &8.0&8.0&8.0 \\
Photon        &8.0&8.0&8.0   &4.0&4.0&4.0   &4.0&4.0&4.0 \\
Kine. fit.      &4.0&4.0&4.0   &4.0&4.0&4.0   &4.0&4.0&4.0 \\
Bg. contam.       &0.0&7.0&0.0   &0.0&0.0&0.0   &0.0&0.0&0.0 \\
Bg. shape      &13.6&7.3&11.6 &9.6&4.5&6.3   &0.0&13.3&0.0 \\
Luminosity &4.0&4.0&3.0   &4.0&4.0&3.0   &4.0&4.0&3.0 \\ 
$B(X \rightarrow Y)$ &0.8&0.8&0.8   &0.7&0.7&0.7   &3.4&3.4&3.4  \\ 
MC fluct.      &1.4&1.4&1.4   &1.3&1.3&1.3   &2.1&2.1&2.1 \\ 
Generator &  1.3&1.3&1.3 & 3.4&3.4&3.4 & 2.8&2.8&2.8     \\ 
$\mathcal{F}_{VP}(s)$ &2.5&2.5&2.5  &2.5&2.5&2.5   &2.5&2.5&2.5 \\ \hline
Sum $(\sigma^{sys})$ & 17.5&15.0&15.8  & 13.3&10.2&10.8  &11.9&17.9&11.6 \\
\end{tabular}
\end{ruledtabular}
\end{table}
\section{Results and discussion}
The Born order cross sections for $\eeto X$
are calculated from
\begin{equation}\label{cross}
\sigma_{Born} = \frac{ N^{obs}_{\eeto X\rightarrow Y}}
{B(X\rightarrow Y)\mathcal{L} \cdot \epsilon^{MC} \cdot (1+\delta)},
\end{equation}
where $N^{obs}_{\eeto X\rightarrow Y}$ is the number of observed events
 in the final state from one photon annihilation, 
 X is the intermediate state, Y is the final state, 
 $\epsilon$ is the detection efficiency obtained from the MC
simulation,  $\mathcal{L}$ is the integrated luminosity,
 and $\delta$ is the radiative correction calculated according to 
Ref.~\cite{gbfm}. With the cross section of $\eeto VP~(\vp)$ from
 Eq. (\ref{cross}) and using 
Eq. (\ref{Fzh}), we obtain $|\mathcal{F}_{VP}|$ 
at $\sqrt{s}=$3.650, 3.686, and 3.773 GeV.

The branching fraction for $\psi(2S)\rightarrow X$ is calculated from
\begin{equation}
B(\psi(2S)\rightarrow X)
  =\frac{N^{obs}_{\psi(2S)\rightarrow X\rightarrow Y}}
   {N_{\psi(2S)}\cdot B(X\rightarrow Y)\cdot \epsilon^{MC}},
\end{equation}
where $N^{obs}_{\psi(2S)\rightarrow X\rightarrow Y}$ is the number of 
observed events in the final state from $\psip$ decay.

Tables~\ref{Teevp} and \ref{Tpsip} summarize the results for $\eetowpi$, 
$\rho\eta$, and $\rho\etap$ at $\sqrt{s}=$3.650, 3.686, and 3.773 GeV, and the
branching fractions of $\psipto\wpi$, $\rho\eta$, and $\rho\etap$ at
3.686  GeV. 
The ratios of $\psip$ to $\jpsi$ 
branching fractions  are also
listed in Table~\ref{Tpsip}, where the $\jpsi$ branching fractions are taken
from the PDG~\cite{PDG}. 

\begin{table*}
\caption{\label{Teevp} Cross sections and form factors measured for
$\eeto\wpi$, $\rho\eta$, and $\rho\etap$ at $\sqrt{s}=3.650$, 3.686, 
and 3.773 GeV. }
\begin{ruledtabular}
\begin{tabular}{l|c|c|c|c|c|c|c} 
Channel & Samples  & $\mathcal{L}$~(pb$^{-1}$)  & $N^{obs}_{Cont.}$ 
& $\epsilon$ (\%) & $1+\delta$ & $\sigma_{0}$ (pb) & 
                           $|\mathcal{F}_{VP}|(\GeV^{-1})$ \\ \hline
       & 3.650 GeV    & 6.42    & $7.3^{+3.3}_{-2.7}$     &  5.09  & 1.032   
          &  $24.3^{+11.0}_{-9.0}\pm4.3$    &  $0.051^{+0.12}_{-0.10}$   \\  
$\wpi$ & 3.686 GeV    & 19.72   & $17.3^{+5.7}_{-5.1}$    &  4.98  & 1.031   
          & $19.2^{+6.3}_{-5.7}\pm2.9$     &  $0.045^{+0.008}_{-0.007}$   \\  
       & 3.773 GeV    & 17.3    & $8.6^{+4.0}_{-3.3}$  &  5.09     & 1.028
       & $10.7^{+5.0}_{-4.1}\pm1.7$     &  $0.034^{+0.008}_{-0.007}$ \\ \hline
          & 3.650 GeV    & 6.42    & $2.3^{+2.1}_{-1.4}$    &  10.9  & 1.028   
          &  $8.1^{+7.4}_{-4.9}\pm1.1$    &  $0.030^{+0.014}_{-0.009}$   \\  
$\rho\eta$ & 3.686 GeV    & 19.72   & $16.0^{+5.6}_{-5.0}$   &  10.9  & 1.028  
           & $18.4^{+8.6}_{-7.8}\pm1.9$     &  $0.046^{+0.011}_{-0.010}$   \\  
          & 3.773 GeV    & 17.3    & $5.8^{+3.3}_{-2.6}$    &  10.7  & 1.026
          & $7.8^{+4.4}_{-3.5}\pm0.08$ &  $0.030^{+0.009}_{-0.007}$ \\ \hline
           & 3.650 GeV    & 6.42    & $<4.4$    &  4.33  & 1.021   
               &  $<89$    &  $<0.192$     \\  
$\rho\etap$ & 3.686 GeV    & 19.72   & $2.9^{+2.4}_{-1.6}$  &  4.43  & 1.020   
           & $18.6^{+15.4}_{-10.3}\pm3.6$  &  $0.050^{+0.021}_{-0.015}$   \\  
           & 3.773 GeV    & 17.3    & $<3.9$     &  4.56  & 1.019
              & $<28$   &  $<0.106$    \\  
\end{tabular} \\
\end{ruledtabular}
\end{table*}
\begin{table*}
\caption{\label{Tpsip} Branching fractions measured for $\psipto\wpi$, 
$\rho\eta$, and $\rho\etap$. The  corresponding
$J/\psi$ branching fractions~\cite{PDG} and the
ratios $Q_h=\frac{B(\psi(2S)\rar h)}{B(J/\psi)\rar h)}$ are also given.}
\begin{ruledtabular}
\begin{tabular}{l|c|c|c|c|c|c} 
Channel & $N_{\psip}$ & $N^{obs}_{Res.}$ & $\epsilon$ (\%) 
 & $\calB_{\psipto}~(\times 10^{-5})$ & $\calB_{\jpsito}~(\times 10^{-4})$ 
&  $Q_{h}$ (\%) \\  \hline
$\wpi$ & 1.4$\times 10^{7}$         & $13.9^{+5.1}_{-4.6}$  & 5.97       
   & $1.87^{+0.68}_{-0.62}\pm0.28$  & 4.2$\pm$0.6  
   & $4.4^{+1.8}_{-1.6}$  \\ \hline
$\rho\eta$ & 1.4$\times 10^{7}$     & $13.2^{+5.0}_{-4.6}$  & 13.43 
   & $1.78^{+0.67}_{-0.62}\pm0.17$  & 1.93$\pm$0.23 
   & $9.2^{+3.6}_{-3.3}$ \\ \hline
$\rho\etap$ & 1.4$\times 10^{7}$    & $2.5^{+2.2}_{-1.5}$  & 5.48  
   & $1.87^{+1.64}_{-1.11}\pm0.33$  & 1.05$\pm$0.18 
   & $17.8^{+15.9}_{-11.1}$ \\
\end{tabular} \\
\end{ruledtabular}
\end{table*}
\begin{figure}[hbt]
\includegraphics[width=0.45\textwidth]{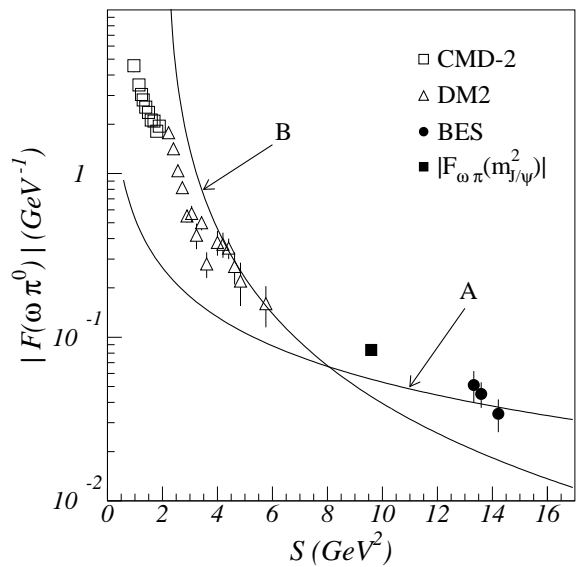}
\caption{\label{Ffwpi} Energy dependence of the $e^+ e^- \to \gamma^*\to\wpi$ form factor. 
Curve (A) is calculated with Eq. (\ref{Formula2}), 
while curve (B) is calculated with Eq. (\ref{Formula1}). }
\end{figure}

Fig. \ref{Ffwpi} shows the measured results of $|\mathcal{F}_{\wpi}|$
from our measurements, CMD-2~\cite{cmd2},  and DM2~\cite{dm2}, and 
the calculated value of $|\mathcal{F}_{\wpi}|$ 
at $s=m^2_{\jpsi}$~\cite{coffman}:
\begin{equation}\label{FFbrwpi}
{\frac{|\mathcal{F}_{\wpi}(m^{2}_{\jpsi})|^2}{|\mathcal{F}_{\wpi}(0)|^2}}=
{\frac{{\frac{\alpha}{3}}({\frac{p_{\gamma}}{p_{\omega}}})^{3} \cdot
m_{\jpsi} \cdot \Gamma(\psip\rar\omega\pi^0)}
{\Gamma({\omega\rar\gamma\pi^0}) \cdot \Gamma(\jpsi\rar\mu^+\mu^-) }}, 
\end{equation} 
where  $|\mathcal{F}_{\wpi}(0)|=2.3 \GeV^{-1}$~\cite{chernyak}, and the other 
quantities are taken from the PDG \cite{PDG}.
Curve (A) is  predicted by
J.-M. G\'{e}rard and G. L\'{o}pez Castro~\cite{Gerard} as:
\begin{equation}
|\mathcal{F}_{\omega\pi^0}(s\to \infty) |
=\frac{f_{\omega} f_{\pi}}{3\sqrt{2}s},
\label{Formula2}
\end{equation}
with the decay constants of  $\omega$ and $\pi$ of $f_{\omega}=17.05\pm0.28$ 
and $f_{\pi}=0.1307 \GeV$. Curve (B) is 
predicted by Victor Chernyak~\cite{chernyak}:
\begin{equation}
|\mathcal{F}_{\omega\pi^0}(s)|  =  |\mathcal{F}_{\omega\pi^0}(0)|
\frac{m^2_{\rho} M^2_{\rho'}}{(m^2_{\rho}- s)(M^2_{\rho'} - s)},
\label{Formula1}
\end{equation}
where $m_{\rho}$ and $M_{\rho'}$ are the masses of $\rho(770)$ 
and $\rho(1450)$, respectively. 
From  Fig. \ref{Ffwpi}, our results agree with
the description of Eq. (\ref{Formula2}).

\section{SUMMARY}
In conclusion, we determine branching fractions  for
$\psiptoomegapi$, $\rho\eta$, and $\rho\etap$ and the 
form factors $\mathcal{F}_{\wpi}$, $\mathcal{F}_{\rho\eta}$, and 
$\mathcal{F}_{\rho\etap}$ at $\sqrt{s}=$3.650, 3.686,
and 3.773 GeV for the first time. 
The branching fractions of $\psipto\wpi$ and $\rho\eta$ 
in our measurement are consistent with those of CLEOc~\cite{cleo}.

\acknowledgments
   The BES collaboration thanks the staff of BEPC for their hard efforts.
This work is supported in part by the National Natural Science Foundation
of China under contracts Nos. 19991480, 10225524, 10225525, the Chinese Academy 
of Sciences under contract No. KJ 95T-03, the 100 Talents Program of CAS 
under Contract Nos. U-11, U-24, U-25, and the Knowledge Innovation Project of 
CAS under Contract Nos. U-602, U-34 (IHEP); by the National Natural Science 
Foundation of China under Contract No. 10175060 (USTC), and No. 10225522 (Tsinghua University); 
and by the Department 
of Energy under Contract No. DE-FG03-94ER40833 (U Hawaii).


\begin{thebibliography}{99}
\bibitem{howl-vp}In the generator for  $\psip$ decays into
VP mesons, the angular distribution is described by 
${\frac{d^3\sigma}{d\cos\theta_V d\cos\theta_1 d\phi_1}}$
$={\sin^2\theta_1[1+\cos^2\theta_V+\sin^2\theta_V\cos(2\phi_1)]},$
where $\theta_V$ is the angle between the vector meson and the positron
direction. For two-body decays, such as $\rho\rar{\pipi}$, $\theta_1$ and
$\phi_1$ are the polar and azimuthal angles of the momentum of the 
$\pi^+$ with respect to the helicity direction of the $\rho$.
For $\omega\rar\pipipi$, $\theta_1$ and
$\phi_1$ represent the angles of the normal to the decay plane with respect
to the helicity direction of  the $\omega$.
\bibitem{Chisquare}
 The combined $\chi^2$, $\chi_{com}^{2}$, is defined as the sum
of the $\chi^2$ values of the kinematic fit ($\chi^{2}_{kine}$) and those from
each of all particle identification assignments: 
$\chi_{com}^{2}=\sum_{i}\chi^{2}_{PID}(i)+\chi^{2}_{kine}.$
\bibitem{significance}
  The significance $S$ is calculated by $S=[-2\times 
(lnL_2-lnL_1)]^{1/2}$, 
where $L_1$ and $L_2$ is the likelihood function value in the fit with and 
without signal, respectively.
\end{thebibliography}
\end{document}